# LLM-based Frameworks for Power Engineering from Routine to Novel Tasks

**Ran Li[1,2,3], Chuanqing Pu[3], Junyi Tao[1], Canbing Li[1,3], Feilong Fan[3]✉, Yue Xiang[4], Sijie Chen[1,3]**

**The digitalization of energy sectors has expanded the coding responsibilities for power engineers and researchers. This research article explores the potential of leveraging Large Language Models (LLMs) to alleviate this burden. Here, we propose LLM-based frameworks for different programming tasks in power systems. For well-defined and routine tasks like the classic unit commitment (UC) problem, we deploy an end-to-end framework to systematically assesses four leading LLMs—ChatGPT 3.5, ChatGPT 4.0, Claude and Google Bard in terms of success rate, consistency, and robustness. For complex tasks with limited prior knowledge, we propose a human-in-the-loop framework to enable engineers and LLMs to collaboratively solve the problem through interactive-learning of method recommendation, problem decomposition, subtask programming and synthesis. Through a comparative study between two frameworks, we find that human-in-the-loop features like web access, problem decomposition with field knowledge and human-assisted code synthesis are essential as LLMs currently still fall short in acquiring cutting-edge and domain-specific knowledge to complete a holistic problem-solving project.**

The digitalization of energy systems has led to a boost in coding and programming tasks such as forecasting[1,2], scheduling[3,4] and control[5,6]. The transition from traditional rule-based to program-based approaches has encompassed a wide range of applications, including tasks such as UC[7,8] and large-scale renewable scheduling[9,10]. As a result, power engineers are now drowning in various programming languages and software tools and are facing two challenges. First, routine tasks such as daily UC and real-time economic dispatch demand repetitive coding and debugging processes, which can be time-consuming. This involves accommodating different locations, rolling timeframes and different constraints. Second, for unexplored problems such as ultra-fast renewable scheduling, engineers must be adept at organizing technology roadmaps and developing a deep understanding of problem modeling and coding, thereby requiring versatility in prior knowledge.

The recent development of LLMs like ChatGPT has shown the potential to relieve the burden. It is demonstrated that beyond its mastery of language, LLM can solve novel and difficult tasks that span mathematics, coding, vision, medicine, law, psychology and more[11]. Users can describe their requirements in natural language and achieve semi-automatic or fully automatic modeling and coding, enabling engineers and researchers to focus on domain-specific problem-solving and design[12,13]. In the area of mathematics and physics, researchers have proved that LLMs can program numerical algorithms[14]. In the area of chemistry, LLMs can generate functional computer code related to chemical equations, chemical structures, units and principles[15]. In the area of electronic engineering, researchers have utilized LLMs to assist in the writing of Hardware Description Language (HDL)[16].

Inspired by successful applications in other sectors, this paper explores the question of whether power engineers could similarly benefit from such advances and in what capacity. Despite an extensive literature review, we discovered no reports concerning the application of LLM-based models within the energy domain. To fill this research gap, we propose two LLM-based frameworks energy system applications as illustrated in Fig. 1, ranging from routine tasks to innovative tasks. For routine tasks, we adopt a straight forward end-to-end framework to evaluate four LLMs including ChatGPT 3.5, ChatGPT 4.0, Claude and Google Bard in terms of success rate, consistency and robustness. The UC problem is employed as a representative example. For innovative tasks, we propose a human-in-the-loop framework including method recommendation, problem decomposition, subtask programming and synthesis. The problem of accelerating large-scale power system dispatching is tested. The frameworks enable power engineers and researchers to select appropriate prompts and steps that fully utilize the LLMs hinging upon the problem's categorization.

[1]Key Laboratory of Control of Power Transmission and Conversion, Ministry of Education, and Shanghai Non-Carbon Energy Conversion and Utilization Institute, Shanghai Jiao Tong University, Shanghai 200240, China. [2]Department of Electronic and Electrical Engineering, University of Bath, UK. [3]College of Smart Energy, Shanghai Jiao Tong University, Shanghai, China. [4]College of Electrical Engineering, Sichuan University, Chengdu, China. ✉ e_mail: feilong-fan@sjtu.edu.cn.

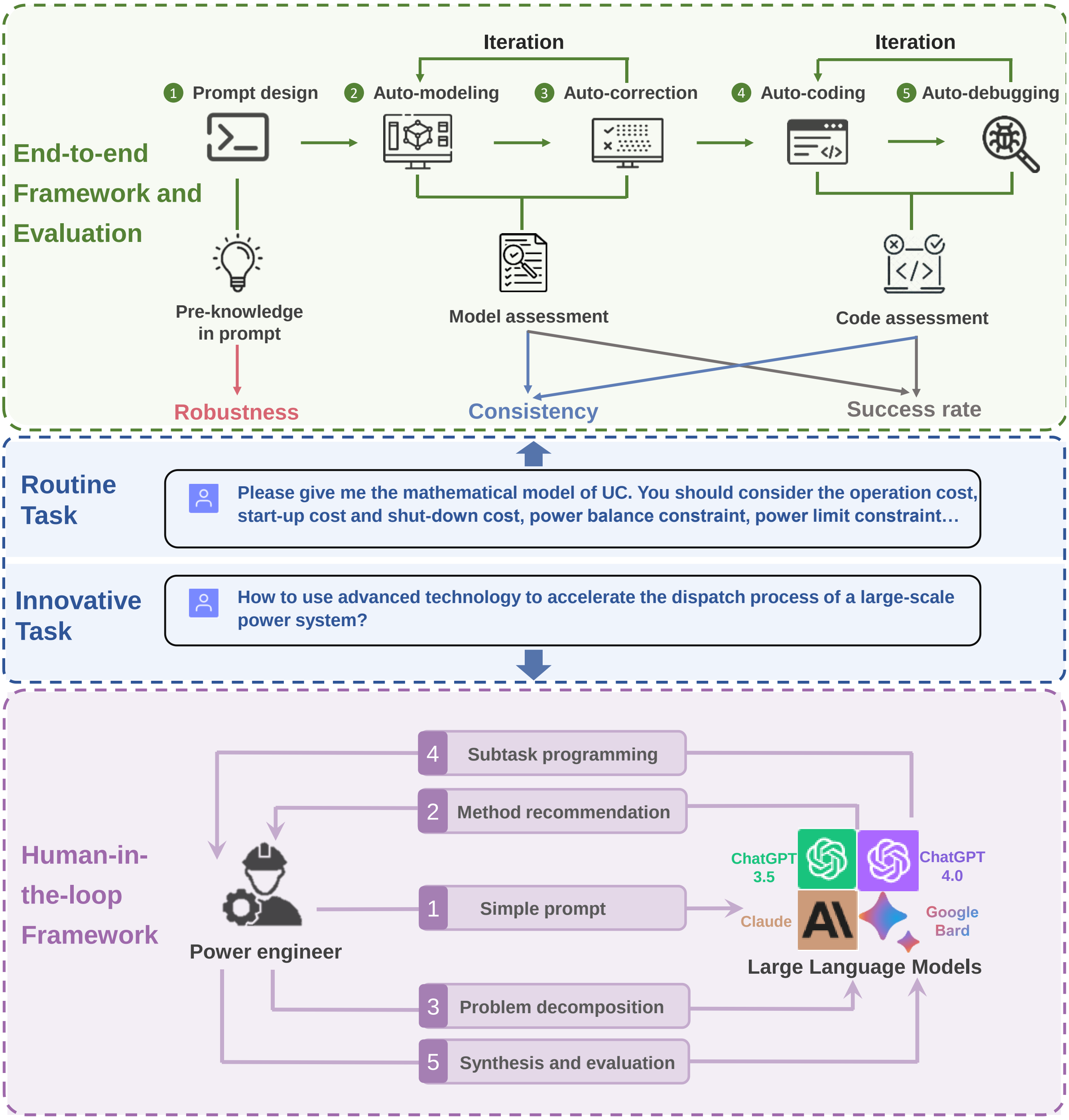

**Fig. 1 | Two LLM-based frameworks for different programming tasks in power systems.** This figure proposes two frameworks for power engineers to solve programming tasks in the energy domain. **a**, an end-to-end framework for LLMs to address routine power system tasks where power engineers know the solution but need time-saving automation. The framework includes: 1 prompt design, 2 auto-modeling, 3 auto-correction, 4 auto-coding and 5 auto-debugging. A set of evaluation metrics are designed to assess LLMs on a multi-metric scale, encompassing pre-knowledge in prompt, model assessment and code assessment metrics in terms of success rate, consistency, and robustness. **b**, an human-in-the-loop framework to empower LLMs to solve innovative tasks in power systems where power engineers have limited prior knowledge and can only describe it in simple and abstract natural language. The framework includes: 1 simple prompt, 2 method recommendation, 3 problem decomposition, 4 subtask programming and 5 synthesis and evaluation. Features like web access, problem decomposition with field knowledge and human-assisted code synthesis are helpful for the successful completion of innovative tasks.

## Solving routine tasks with the End-to-end Framework

We evaluate the performance of 4 LLMs using unit commitment, which is a classic routine task for power system operation, involving the determination of an optimal schedule for generating units to meet the expected demand over a given time horizon while minimizing costs and adhering to other operational generation constraints. UC is typically formulated as a mixed-integer linear programming (MILP) problem. The task can be time-consuming involving repetitive coding and debugging process. LLM could automate the work for engineers through the end-to-end framework including prompt design, auto-modeling, auto-correction, auto-coding and auto-debugging. Typical modeling and coding processes are exemplified in Fig. 2.

Although the end-to-end framework is straightforward, the focus here is to evaluate performances of different LLMs on routine tasks. For modelling tasks, we would identify whether the model is correct using engineering experience and then point out the error type to the LLM through new prompts without specifically indicating how to modify it. For coding tasks, we first ran the code provided by the LLM to pinpoint bugs. If there was a bug, the error message from the language environment (e.g. Python interpreter) was sent directly back to the LLM for modification. If the program ran successfully, we finally checked whether a UC solution matched the optimal solution produced by commercial power system dispatching software. If inconsistencies were found, this information was directly fed back to the LLM for modification. With such iterations between power engineers and LLMs, it is possible for LLMs to automatically program the UC problem. The detailed process is exemplified in Supplementary from Table 1 to Table 4.

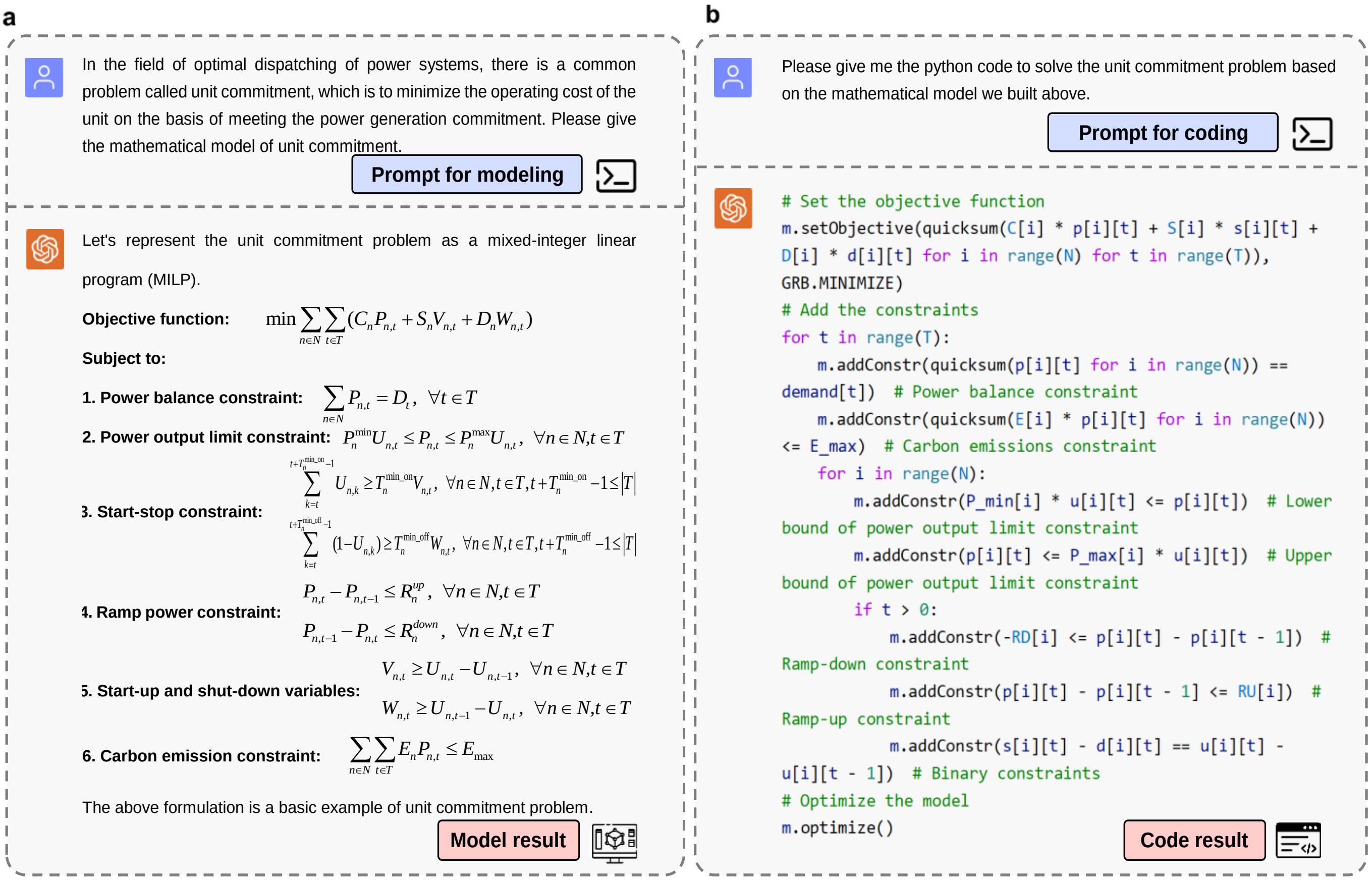

**Fig. 2 | Auto-modeling and auto-coding of LLM for the UC problem.** This figure demonstrates the basic modeling and coding process of LLM under the prompt of researchers. **a**, in response to the prompt for modeling, LLM gave out the basic mathematical model of UC, including an objective function for minimizing the total cost of generators and generation constraints such as power balance constraint and power output limit constraint. **b**, in response to the prompt for coding, LLM used Gurobi optimizer as a solver and successfully wrote the Python code to solve the UC problem based on the mathematical model built above.

## Evaluation Metrics

**Success rate:** The metric of *success rate* refers to the rate of successful modeling or coding tasks completed in repeated experiments shown in Fig. 3. At the model level, we focus on objective correctness, constraints correctness and completeness. 'Objective correctness' assesses if the objective function holistically includes the fuel cost, start-up and shut-down costs across all scheduling periods and generation units. 'Constraints correctness' ensures constraints are accurately defined in relation to the decision variables and given parameters, and 'Constraints completeness' verifies that constraints envelop the entire spectrum of requirements like unit power range, start-stop relationship transformation, and any bespoke constraints from user prompts. At the code level, we first use ‘Error-free execution’ to test if the code generated by the LLM can execute bug-free. On top of that, we further compare the dispatching decisions from LLM’s code with those from commercial software using ‘Decision verification’.

**Robustness:** Since different users have unequal pre-knowledge, their prompts vary in quality and may affect the performance of LLMs. To account for this variability, we have designed a range of prompts, from basic to sophisticated. The metric *robustness* refers to the success rate of the LLM under the worst-case scenario which is the basic prompt in Fig. 3. Examples of these can be found in Supplementary Table 5, where simple prompt might ask ‘*Please give me the mathematical model of UC*’ while more sophisticated ones may ask ‘*Please give me the mathematical model of UC. You should consider the operation cost, start-up cost and shut-down cost of generators in the objective. The constraints include power balance constraint, power limit constraint, carbon emission constraint…*’.

**Consistency:** A notable challenge with LLMs is non-replicability, that is, they might produce different outputs for the same input due to the inherent stochastic nature of the model. The metric *consistency* reflects whether the results given by LLMs can be reliably reproduced regardless of their correctness. While it's challenging to obtain word-for-word identical answers, we consider a response consistent if it fulfils the same function. As shown in Fig. 3, this metric corresponds to the number of times the output is repeated across three independent trials.

For this research, we designed three different levels of pre-knowledge in prompts and each prompt is repeated three times. Fig. 3 exemplifies this using ChatGPT 3.5. The numbers in brackets refer to the iterations required to achieve a satisfactory answer through human feedback, that is, the number of prompts required for auto-correction and auto-debugging. Typical feedback prompts in iteration process are shown in Supplementary Table 6.

| Prompt type | Repeated trials | Model assessment | | | | Code assessment | |
|---|---|---|---|---|---|---|---|
| | | Objective correctness | Constraints correctness | Constraints completeness | Total model Correctness | Error-free execution | Decision verification |
| Simple description | Trial 1 | √(3) | √(1) | √(1) | √(3) | √(0) | √(3) |
| | Trial 2 | √(0) | ×(3) | √(2) | ×(3) | √(0) | ×(3) |
| | Trial 3 | √(0) | ×(3) | √(3) | ×(3) | √(1) | ×(3) |
| Intermediate description | Trial 1 | √(0) | √(1) | √(0) | √(1) | √(3) | √(0) |
| | Trial 2 | √(0) | ×(3) | √(0) | ×(3) | √(1) | ×(3) |
| | Trial 3 | √(3) | √(1) | √(1) | √(3) | √(1) | √(3) |
| Sophisticated description | Trial 1 | √(0) | √(2) | √(0) | √(2) | √(2) | √(0) |
| | Trial 2 | √(0) | √(0) | √(0) | √(0) | √(1) | √(2) |
| | Trial 3 | √(0) | √(0) | √(1) | √(1) | √(1) | √(2) |

Robustness = 1 (Total model Correctness); Robustness = 1 (Decision verification); Consistency = 2, Success rate = 2 (Total model Correctness); Consistency = 3, Success rate = 3 (Decision verification)

**Fig. 3 | Evaluating ChatGPT 3.5 in the UC task.** This figure demonstrates the evaluation results of ChatGPT 3.5. In the far left column under the 'prompt type' section, we differentiated three types of prompts: from simple description to sophisticated description. We conducted three independent, repeated trials for the same category of prompts. We employed a straightforward tick ‘√’ notation in the corresponding table if the result is accurate and the number of iterations is three or fewer. Conversely, if these criteria aren't met, an ‘×’ is marked. It's important to note that a ‘√’ is ascribed to the overall model correctness metric only when all underlying indicators are ticked with ‘√’. If even one fails to meet the mark, the overall indicator is labelled as ‘×’. We accordingly compute metrics such as robustness, consistency, and success rate. The specific calculation methods can be found in the ‘method’ section.

We used these metrics to perform a comprehensive assessment of these LLMs as shown in Fig. 4. The paid subscription model, ChatGPT 4.0 outperforms its counterparts consistently across all evaluation metrics.

Among free LLMs, Google Bard failed in both modeling and coding. This may be associated with a lack of exposure to power system issues in its training data. ChatGPT 3.5 and Claude exhibited moderate success rates (2.0 score), showing their ability in programming. Closer error analysis revealed ChatGPT 3.5 and Claude repeatedly failed to model accurate minimum start-up and shut-down time constraints, especially under the prompts with 'Simple description'.

In terms of robustness, free LLMs displayed noticeable deficits under low-quality prompts. It reflects the lack of expertise in the power and energy sector. For example, they may struggled with the representation of start-up and shut-down costs, which involve the introduction of integer variables and the linearization of the model—aspects not hinted at in the simple prompts. In conclusion, users with limited prior knowledge will benefit most from paid models like ChatGPT 4.0. Among free models, ChatGPT 3.5 has the possibility to successfully complete the task under the simple prompt, achieving relatively higher scores on robustness than Claude. It shows the ChatGPT 3.5 model may have higher inherent variability in their architecture and training data.

This variability, on the other hand, causes ChatGPT 3.5 scoring lower on consistency compared with Claude. As this inconsistency only occurs under the simple prompt, one possible reason is the ambiguity of the prompt triggers the model to explore different styles of answers. Other LLMs generally scored well on the consistency metric, indicating that they are capable of replicating power system programs given the same prompts.

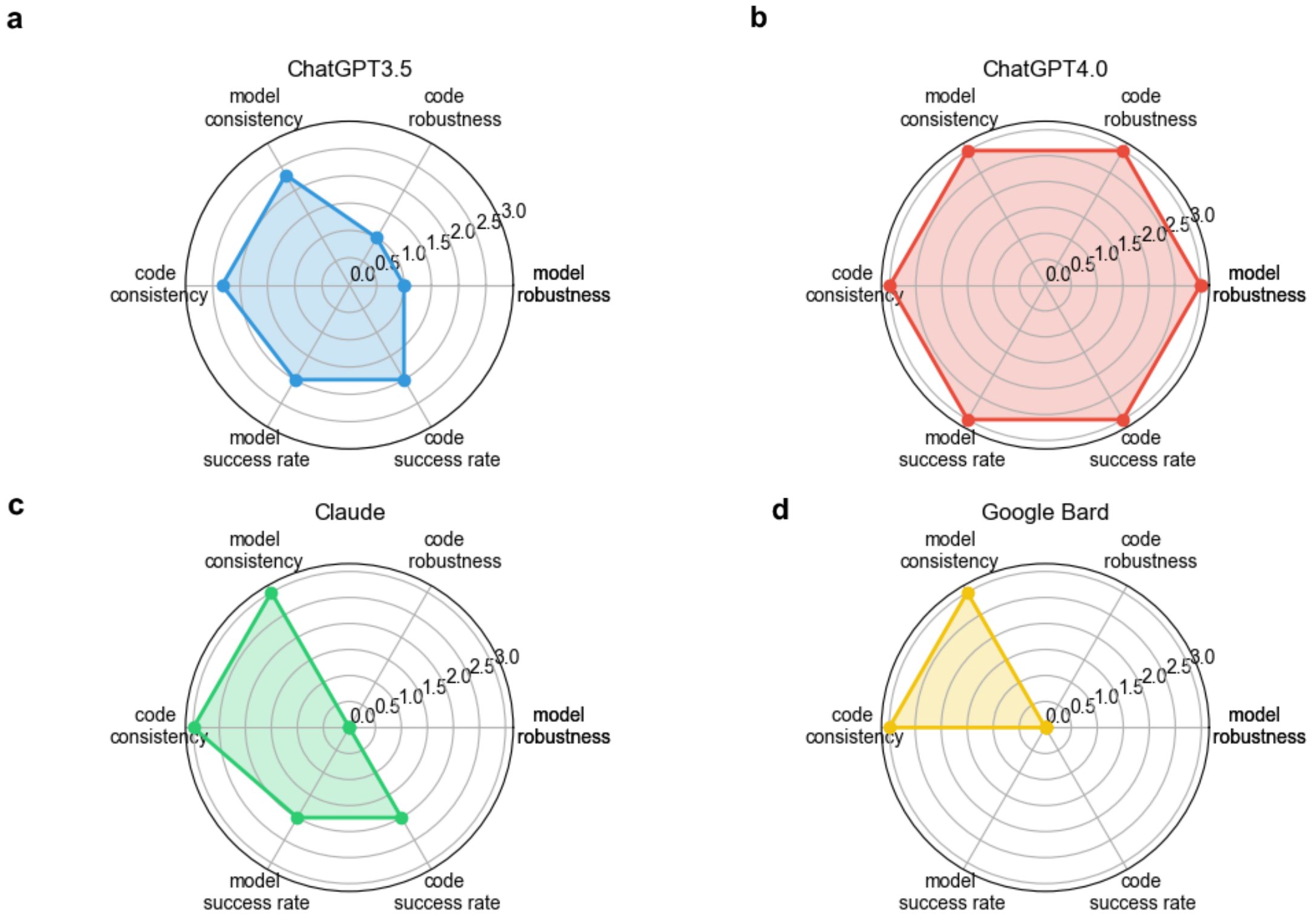

**Fig. 4 | Comparative analysis of LLMs in addressing UC.** This figure quantifies the proficiency of various LLMs including ChatGPT 3.5, ChatGPT 4.0, Claude and Google Bard in handling UC tasks, as assessed through our devised six metrics: 'model consistency', 'code consistency', 'model success rate', 'code success rate', 'model robustness' and 'code robustness'. This graphical representation comprises of six axes, each corresponding to one of our evaluation metrics. Each axis spans from the centre to the outer edge of the radar, with the outer edge indicating superior performance and the centre indicating lesser performance. By interconnecting the values along these axes for each LLM, the radar chart reveals the multifaceted capabilities and comparative strengths and weaknesses in their approach to UC modeling and coding tasks.

It is difficult for LLMs to produce accurate outputs with a single prompt. Instead, we anticipate LLMs to refine their outputs based on human feedback. Given this approach, it becomes essential to evaluate the efficiency of LLMs in assimilating feedback and rectifying errors across iterations. Additionally, different auto-correction abilities reveal some models may be more receptive and adaptive to feedback than others.

Therefore, we further explored the adaptive capabilities of LLMs via an iteration correctness test. Fig. 5 illustrates the success rates of these LLMs under different feedback iteration times, spanning six scenarios that include three prompt qualities and two tasks: modeling and coding. ChatGPT 4.0 required minimal feedback (no more than one iteration) to correct its responses under all six scenarios while Google Bard failed to deliver satisfactory results after three iterations. Under simple prompts, only ChatGPT 4.0 were able to achieve a full model success rate, with ChatGPT 3.5 at 33.3% and Google Bard and Claude at 0%. As the prompts progressed to sophisticated, the number of iterations required decreased. The initial success rate rose from 0% to finally 66.7% (ChatGPT 4.0), 100% (Claude) and 33.3% (ChatGPT 3.5) in modeling tasks, and 100% (ChatGPT 4.0, Claude) and 33.3% (ChatGPT 3.5) in coding tasks. An interesting observation is while ChatGPT 3.5 and Claude yield similar results in terms of success rate, Claude generally reached optimal accuracy in fewer iterations in intermediate and sophisticated prompt than ChatGPT 3.5, displaying better learning rates.

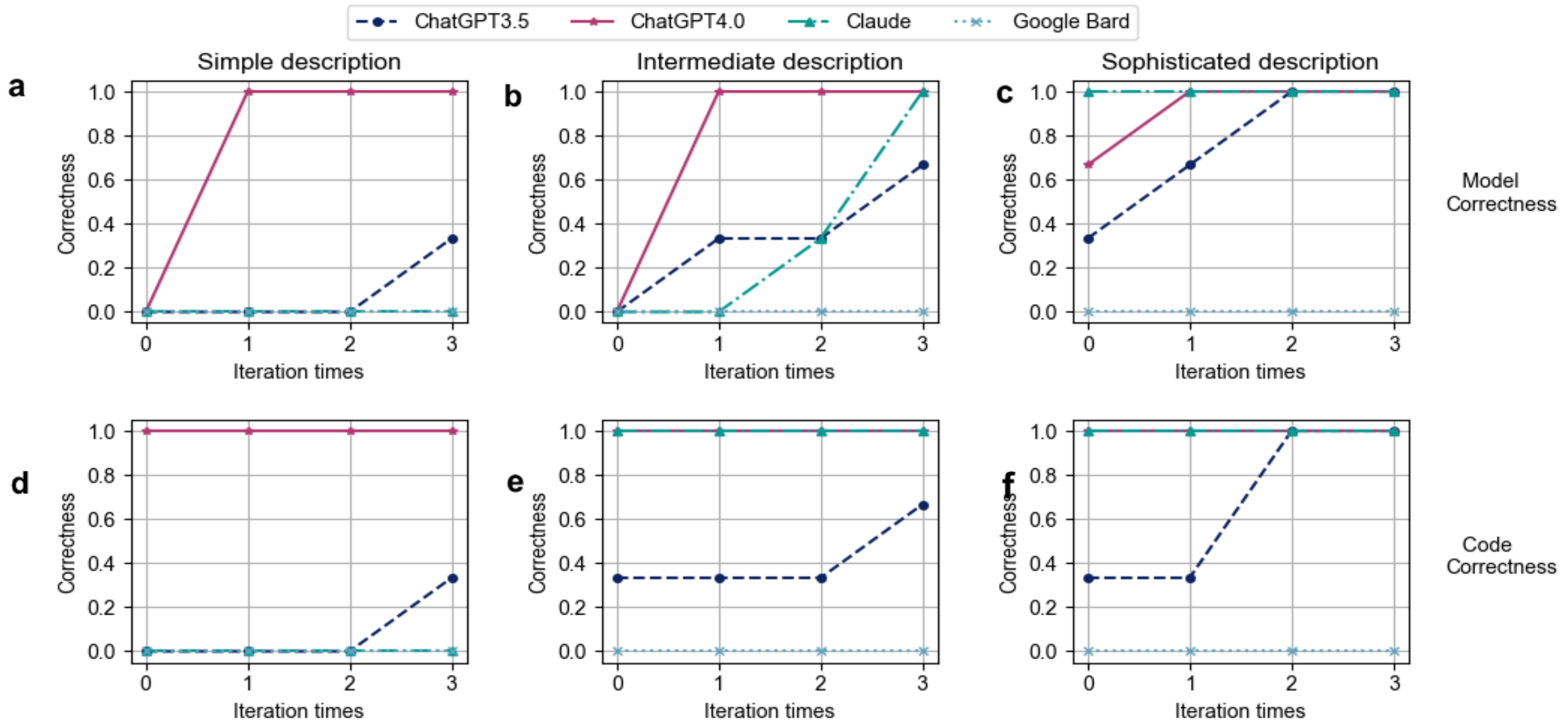

**Fig. 5 | Evaluation of correctness in response to feedback iteration.** The figure is organized as a grid of line charts, distributed across two rows and three columns. Each row (sub-figure **a** to **c**, **d** to **f**) represents the test results under distinct task types—modeling and coding respectively, while each column (sub-figure **a** to **d**, **b** to **e**, **c** to **f**) demonstrates results under different prompt conditions. Every sub-figure within the grid shows the progression of correctness rates for the various LLMs with increasing counts of manual corrections. The correctness metric is deemed successful only when all subtasks under the given task have been successfully completed in a single testing instance. This exposition underscores the LLMs' in-depth exploration of their capacity for error correction and the refinement of their responses in a feedback-driven environment.

## Exploring innovative tasks with LLM – Human-in-the-loop Framework

Building upon the understanding of LLM's capabilities for routine tasks in energy systems, this stage extends to complex applications where engineers or researchers have limited prior knowledge. With the fast development of the smart grid, innovative problems without clear specific solutions keep appearing. These problems lack analytical expression and often come up as abstract questions in natural language. An example of such problems is the acceleration of large-scale power system optimization, especially when considering the increasing uncertainty from renewables. Without much prior knowledge, engineers may start the conversation with a simple prompt such as 'how to use advanced technology to accelerate the dispatch process of a large-scale power system'. It would be difficult for an LLM to fulfil the holistic project from modeling to coding based on this abstract prompt.

Hence, we propose a human-in-the-loop framework to enable LLMs and power engineers to collaboratively address the innovative problem. As shown in Fig. 1 (b), the framework aims to leverage LLM's versatile ability and act as an algorithmic engineer to support the project. We designed the experiment with the best model so far (ChatGPT 4.0) to enable a more focused examination. Upon initiating our dialogue shown in Supplementary Table 7, it is noticed that machine learning methods and advanced optimization algorithms can be used for UC acceleration. Taking machine learning as an example, ChatGPT recommended the following paper 'Solving mixed integer programs using neural networks' and the corresponding method of neural diving (ND) to achieve the acceleration of UC in large-scale power systems. It may not be an easy task for power engineers to develop a new program for this specific application from scratch. Our idea is to leverage ChatGPT to accomplish this through step-by-step questioning.

Following the framework, ChatGPT broke the problem down into 6 steps: 1. Data Collection and Pre-processing, 2. Training Data Generation, 3. Neural Network Design and Training, 4. Integration into Branch-and-Bound, 5. Evaluation and Fine-tuning and 6. Deployment, as shown in Supplementary Table 8. As the breakdown is quite broad, we need to further decompose each part into programmable subtasks. This process requires field knowledge in power systems and programming, in which human involvement is beneficial.

After reading the paper, we understand the fundamental idea of ND is to establish a mapping from external conditions (such as load, wind/solar output, and generation cost) to unit status (on/off) through a graph convolution neural network (GCNN). By directly predicting the on/off status in UC, integer variables are constrained, transforming the original MILP issue into a Linear Programming (LP) problem, thus achieving UC acceleration. Based on this knowledge, we manually designed the technical roadmap as depicted in Extended Data Fig. 1.

So far, power engineers have decomposed it into six coding subtasks, where each can be automatically completed by the LLM. For example, the first one is 'Data Collection and Pre-processing', which consist of bipartite graph representations of MILP problem instances with varying parameters and their corresponding solutions. Apart from the research paper on ND, no existing tutorials or online codes were found in current search engines on how to convert a MILP instance into a bipartite graph data structure. Additionally, there are no established Python libraries or functions available that can directly read an LP file and convert it into a bipartite graph structure. To our surprise, through its strong semantic understanding, ChatGPT offered researchers valuable programming insights and generated the requisite code to bridge this gap, as delineated in Supplementary Table 9.

Based on the success of 'Data collection and Pre-processing', we also discovered that ChatGPT has the ability to construct neural network models tailored to specific input-output structures for GCNN, which is shown in Supplementary Table 10. Furthermore, it can assist researchers in quickly discovering relevant Python packages for their specific needs. The package-based code provided by ChatGPT significantly reduces the development difficulty for engineers. Detailed prompts, responses and codes for neural diving training are presented in Supplementary Table 11.

Conclusively, LLM can quickly search and recommend appropriate papers and methods to kick off the project. After learning the overall technical roadmap, power engineers need to manually decompose it into programmable subtasks using field knowledge. For a well-decomposed problem, each subtask is equivalent to the routine tasks described in the first part of the paper. We can utilize the end-to-end framework to model the subtasks analytically and program corresponding bug-free codes. These subtasks need to undergo code synthesis to achieve full functionality and be integrated into a unified project. Power engineers needed to leverage their expertise to synthesize key sub-modules, and also make necessary adjustments to the code, such as changes in file paths and tuning of model hyperparameters to ensure optimal project performance and cohesiveness.

**Comparative Study.** One of our key finding is that human-in-the-loop is still required for power system innovative tasks. We will show this through a comparative study between two frameworks on the same task. The interactive learning between human and LLM is illustrated on the left side of Fig. 6. First, it's noted that due to the nature of language generation models, it cannot guarantee that LLM always provides 'real' literature. Also, since the training data of ChatGPT is stuck in 2022, it cannot ensure to provide the latest results. Here utilize the 'WebChatGPT', a plugin designed to facilitate the web-enabled environment, where ChatGPT referenced an authentic paper 'Solving mixed integer programs using neural networks' in arXiv in 2020. Power engineers can download and learn key methodologies in this paper. In turn, engineers could enhance its prompts from abstract to specific by detailing the subtasks like forging optimal joint variable assignments and narrowing the objective function variances. It forms an interactive learning process between humans and LLM.

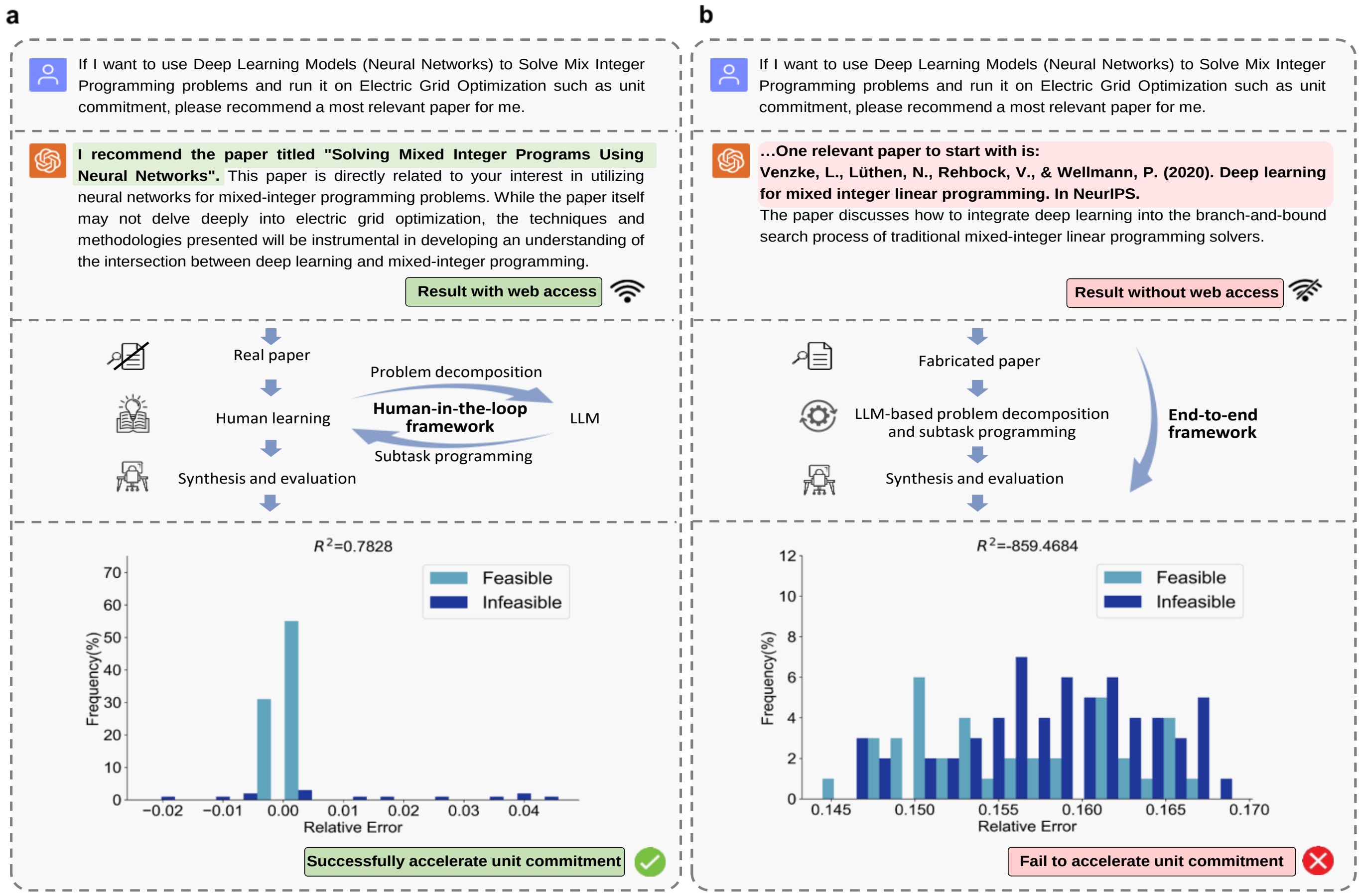

**Fig. 6 | Comparative performance of UC acceleration models under Human-in-the-loop and end-to-end frameworks.** The upper half of the figure shows the recommended methods and literature under two frameworks. Following the methods recommended, the lower half of the figure showcases the predictive accuracy and feasibility (i.e., whether predicted solutions satisfy constraints) of both models across 100 UC samples. The bar plot's x-axis represents the relative error between the operating cost under LLM's decisions and the optimal operating cost using commercial solvers. The y-axis indicates the occurrence frequency of the error interval. Predictions deemed feasible are in light blue while infeasible ones are in dark blue. **a**, panel a depicts results achieved under human-in-the-loop framework. The solutions are mostly feasible with high accuracy. **b**, panel b depicts results achieved using end-to-end framework. The alternative approach predominantly produces infeasible solutions with diminished precision.

The bottom of Fig. 6 (a) showcases the accuracy of our model trained using the human-in-the-loop framework. Out of a test set comprising 100 samples, 86% of the predicted solutions were found to be feasible. The operating cost is close to the optimal, with an $R^2$ value of 0.783. This high accuracy can be attributed to the model's primary function, which is to predict unit status (on/off) for the UC problem. Once these integer solutions are established, the model addresses the remaining subproblems using the physical model containing all constraints, thereby ensuring a high degree of feasibility.

By contrast, Fig. 6 (b) depicted the outcome if we directly employ the end-to-end framework. In the environment without web access, the recommended paper ‘Deep learning for mixed integer linear programming’ in NeurIPS 2020 does not exist. As a result, engineers are unable to learn and feedback high-quality prompts. The rest of this project was solely reliant on the LLM. LLM employed a standard multi-layer perceptron to directly predict all variable values for the UC problem. However, this generic model struggles to simultaneously satisfy all of the constraints for complex power systems. For the same test set, only 39% of the predicted solutions were feasible, and the operating cost deviated from the optimal, resulting in an $R^2$ value of -859.468.

Then, we zoom in the decomposition stage steps to analyse the difference of two frameworks: 1. Engineers read the recommended paper and feedback to ChatGPT through prompts with improved quality for problem decomposition. 2. Ask ChatGPT to perform the problem decomposition autonomously.

Take the first step ‘data collection and pre-processing’ as an example. In Fig. 7 (a), a power engineer outlined the key steps of the subtask, detailed the input and output characteristics, and stressed that the primary objective of dataset generation is to produce bipartite graph data as shown in the green box. Following this prompt, ChatGPT successfully programmed the task and generated the bipartite graph of the dataset.

Conversely, in Fig. 7 (b), ChatGPT was solely tasked with a simple prompt. The step-by-step guide decomposed by ChatGPT is still broad as shown in the red box. The resultant code proved to be bug-free after auto-debugging attempts but only produced an empty dataset which cannot be directly used as training data for the next stage. Thus, while ChatGPT is able to identify relevant papers, it still lacks the ability to digest field knowledge and new techniques.

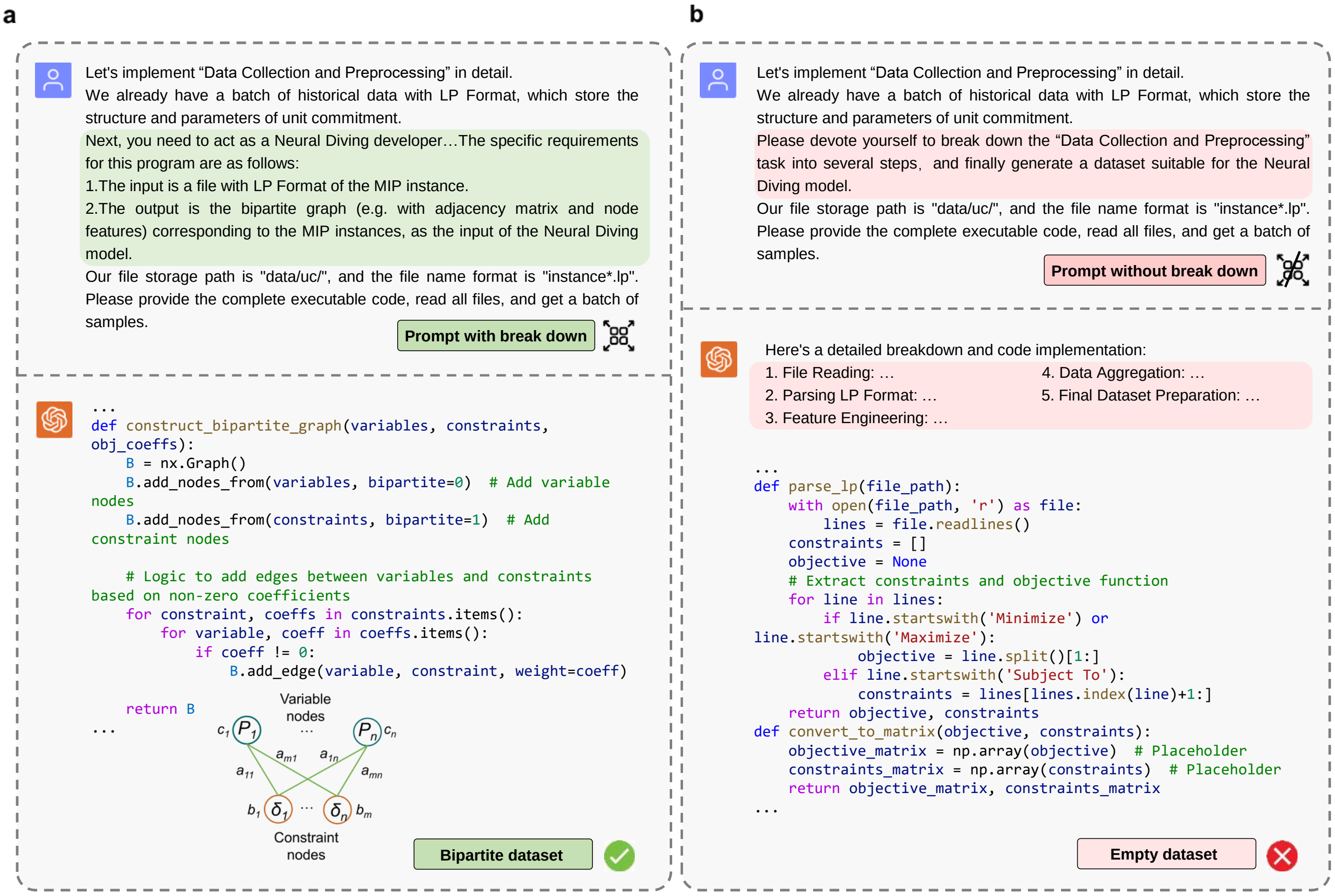

**Fig. 7 | Comparative analysis of code executability for 'Data Collection and Pre-processing' upon different problem decomposition.** **a**, panel (a) depicts the scenario where power engineers specify the input-output requirements for the task: gathering samples from LP-formatted raw data and constructing data in a specific format for subsequent network training (highlighted in green in the left prompt). As a result, LLM-generated code constructs a bipartite graph from LP-formatted UC instances, offering a conducive structure for subsequent neural network learning. **b**, in contrast, panel (b) asks the LLM to break down the task by itself (highlighted in red in the right prompt). The breakdown is generic and the LLM-generated code struggles to effectively extract critical features from LP-formatted raw data, thereby hindering the generation of the subsequent dataset for model training.

## Discussion

This study explores LLMs' ability to assist routine tasks in the energy sector employing a straightforward end-to-end framework that spans from conceptualization to coding and debugging of power system problems. A set of quantitative metrics is designed to compare 4 commercial LLMs in terms of success rate, consistency and robustness when applied to power systems. The results showcase that ChatGPT 4.0 outperforms ChatGPT 3.5, Claude and Google Bard for tasks in power system optimization.

Furthermore, we established a human-in-the-loop framework to capture the capabilities of LLM in addressing innovative challenges within power systems. In this pursuit, we pinpointed key development tips to augment the performance of LLMs. ChatGPT 4.0 demonstrates a notable ability to tackle abstract and unexplored challenges, thereby showcasing its potential as a valuable resource for engineers and researchers addressing the future complexities of low-carbon power systems.

While this primarily exploratory study highlights the potential of the LLM, we also note its limitations. (1) LLMs' databases may be outdated, requiring supplemental measures such as web-based access for updated and authentic information. (2) LLM lacks field knowledge in power knowledge, sometimes resulting in technically infeasible routes. Human-in-the-loop and manual problem-decomposition are still required at this stage. (3) The intrinsic constraint of LLM in retaining code context over prolonged interactions leads to occurrences of the 'memory loss' phenomenon. To mitigate this, human involvement is recommended to synthesize various sub-modules to realise expected functions.

Hence, the present LLMs are more appropriately positioned as algorithmic engineers within the power engineering domain rather than as product managers. While proficient in fundamental tasks, including resource retrieval, methodological integration, model development, and code generation, it falls short in acquiring cutting-edge and domain-specific knowledge to establish technical pathways and maintain a holistic project perspective. This indicates an avenue for future advancements and refinements in LLM to achieve a fully automated power system AI solution.

## Method

**The evaluation metrics.** In the experimental procedure, we established the method for calculating metrics to assess the abilities of LLMs in addressing the UC problem. We instituted a series of binary integer variables $x^{ij}_{m.obj.cor}$, $x^{ij}_{m.con.cor}$, $x^{ij}_{m.con.com}$, $x^{ij}_{c.veri}$, $x^{ij}_{c.bug-free}$ to indicate whether the LLMs achieved certain standards, namely, 'Objective correctness', 'Constraints correctness', 'Constraints completeness', 'Error-free execution' and 'Decision verification'. If these standards were met, the variable was assigned a value of 1; otherwise, it was set to 0. In this context, the superscript index $i$ designates the $i$-th type of prompts, and j denotes the $j$-th trail.

1. The formula for calculating the success rate (SR) is:

$$SR_{\mathrm{m}} = \frac{1}{3}\sum_{i=1}^{3}\sum_{j=1}^{3}\delta(x^{ij}_{m.obj.cor} + x^{ij}_{m.con.cor} + x^{ij}_{m.con.com} = 3) \quad (1)$$

$$SR_c = \frac{1}{3}\sum_{i=1}^{3}\sum_{j=1}^{3}\delta(x^{ij}_{c.veri} + x^{ij}_{c.bug-free} = 2) \quad (2)$$

where $SR_{\mathrm{m}}$ denotes the success rate for modeling, and $SR_c$ signifies the success rate of coding. The function $\delta(\cdot)$ returns a value of 1 if the equation within the parentheses is satisfied; otherwise, it returns 0. Specifically, when $\delta(\cdot)=1$, it indicates that in the $i$-th type of prompt during the $j$-th trial, the LLM successfully completed the modeling or coding task. Summing over three trials yields a success rate metric within the range of 0-3. The outer summation and fraction symbolize the average success rate across the three types of prompts tested.

2. The formula for calculating the consistency (CO) is:

$$CO_{\mathrm{m}} = \frac{1}{3}\sum_{i=1}^{3}\zeta(\sum_{j=1}^{3}\delta(x^{ij}_{m.obj.cor} + x^{ij}_{m.con.cor} + x^{ij}_{m.con.com} = 3)) \quad (3)$$

$$CO_c = \frac{1}{3}\sum_{i=1}^{3}\zeta(\sum_{j=1}^{3}\delta(x^{ij}_{c.veri} + x^{ij}_{c.bug-free} = 2)) \quad (4)$$

where $CO_{\mathrm{m}}$ represents the consistency in modeling, and $CO_c$ signifies the consistency in coding, and $\zeta(\cdot)$ is a piecewise function related to integer variables, as depicted in (5). That is, when all three trials are either successful or unsuccessful, the $\zeta(\cdot)$ returns 3. otherwise, it returns 2. The outer summation and fraction symbolize the average success rate across the three types of prompts tested.

$$\zeta(x) = \begin{cases} 3 & x=0 \\ 2 & x=1 \\ 2 & x=2 \\ 3 & x=3 \end{cases} \quad x \in \mathbb{Z}^{+} \quad (5)$$

3. The formula of the robustness (RO) is:

$$RO_{\mathrm{m}} = \sum_{j=1}^{3}\delta(x^{1j}_{m.obj.cor} + x^{1j}_{m.con.cor} + x^{1j}_{m.con.com} = 3) \quad (6)$$

$$RO_c = \sum_{j=1}^{3}\delta(x^{1j}_{c.veri} + x^{1j}_{c.bug-free} = 2) \quad (7)$$

where $RO_{\mathrm{m}}$ represents the model robustness, and $RO_c$ represents the code robustness. It can be understood that the robustness metric is the success rate when $i$=1 (prompts with simple description).

## Code availability

All the codes used to reproduce these experiments (including the solution code of the UC generated by LLMs, the design code and examples of the ADMM algorithm, and Neural Diving implemented on UC dataset) can be found in https://github.com/BigdogManLuo/ChatGPT-for-Power-System-Programming-Tasks.git.

## Author contributions

R.L conceived the idea. R.L and C.P developed the framework. C.P and J.T developed the codes and undertook the experiments. Y.X and F.F provided the data and LLMs platform. R.L, C.P and J. T. wrote the manuscript. All authors revised the manuscript.

## Competing interests

The authors declare no competing interests.

## Extended data

**1. Links to the dialogue process for the routine task results for the four different LLMs are as follow:**

| LLMs | Prompt type | Links |
|---|---|---|
| ChatGPT 3.5 | Simple description | https://chat.openai.com/share/ca9b2d3b-4325-4fe3-9117-a8d239f69845<br>https://chat.openai.com/share/f52b1395-af3d-499f-aa0b-3e90caa3e186<br>https://chat.openai.com/share/2cbb99e1-7bd1-47ac-83f1-38eb9874eac2 |
| | Intermediate description | https://chat.openai.com/share/4e68bee3-7876-46ce-a907-c9897fffc769<br>https://chat.openai.com/share/7834c60b-f4b2-4b88-a69d-d0d25f42419e<br>https://chat.openai.com/share/5044a655-4801-4506-a5d5-c88ed9285171 |
| | Sophisticated description | https://chat.openai.com/share/fc3409d8-0464-4d15-8441-76c157476f1b<br>https://chat.openai.com/share/5a9f264d-bc09-45ea-82d5-31dd013894de<br>https://chat.openai.com/share/a4ac7315-9823-4dc8-8762-16409d8454ca |
| ChatGPT 4.0 | Simple description | https://chat.openai.com/share/2ac0460d-cba0-4f9b-bb25-1d38695ee656<br>https://chat.openai.com/share/525b6348-00e0-480a-aaaa-cfdbb1784bc9<br>https://chat.openai.com/share/f60bef24-dbd7-496f-b831-ec8110fde78c |
| | Intermediate description | https://chat.openai.com/share/13626372-2626-4d0a-9ab3-817b5ce5b087<br>https://chat.openai.com/share/71f1b70d-e2e6-4677-966e-891028b53c76<br>https://chat.openai.com/share/284e12ce-c41b-4feb-922f-a2a9654efad5 |
| | Sophisticated description | https://chat.openai.com/share/e38ddee7-8f71-40db-b927-16e4c9ec4dcb<br>https://chat.openai.com/share/23617eca-eac2-4e68-8a25-beab487cd7c6<br>https://chat.openai.com/share/607c60dc-0c8d-4b1e-a7e1-ea32447bbcc1 |
| Claude | Simple description | https://github.com/BigdogManLuo/ChatGPT-for-Power-System-Programming-Tasks/blob/master/dialogue/claude/1.1.png<br>https://github.com/BigdogManLuo/ChatGPT-for-Power-System-Programming-Tasks/blob/master/dialogue/claude/1.2.png<br>https://github.com/BigdogManLuo/ChatGPT-for-Power-System-Programming-Tasks/blob/master/dialogue/claude/1.3.png |
| | Intermediate description | https://github.com/BigdogManLuo/ChatGPT-for-Power-System-Programming-Tasks/blob/master/dialogue/claude/2.1.png<br>https://github.com/BigdogManLuo/ChatGPT-for-Power-System-Programming-Tasks/blob/master/dialogue/claude/2.2.png<br>https://github.com/BigdogManLuo/ChatGPT-for-Power-System-Programming-Tasks/blob/master/dialogue/claude/2.3.png |
| | Sophisticated description | https://github.com/BigdogManLuo/ChatGPT-for-Power-System-Programming-Tasks/blob/master/dialogue/claude/3.1.png<br>https://github.com/BigdogManLuo/ChatGPT-for-Power-System-Programming-Tasks/blob/master/dialogue/claude/3.2.png<br>https://github.com/BigdogManLuo/ChatGPT-for-Power-System-Programming-Tasks/blob/master/dialogue/claude/3.3.png |
| Google Bard | Simple description | https://g.co/bard/share/19d51c1d7054<br>https://g.co/bard/share/129fb4eb3e27<br>https://g.co/bard/share/37611eeb2088 |
| | Intermediate description | https://g.co/bard/share/5de33da2e9db<br>https://g.co/bard/share/f190136bbce4<br>https://g.co/bard/share/d944d9fce0a8 |
| | Sophisticated description | https://g.co/bard/share/819dedf35ad6<br>https://g.co/bard/share/bc972d6bc8da<br>https://g.co/bard/share/10266d6288b1 |

## 2. Test results for the four different LLMs are as follow:

Table.A1 Test results of ChatGPT 3.5

| Prompt type | Repeated trials | Model assessment | | | | Code assessment | |
|---|---|---|---|---|---|---|---|
| | | Objective correctness | Constraints correctness | Constraints completeness | Total model Correctness | Error-free execution | Decision verification |
| Simple description | Trial 1 | √(3) | √(1) | √(1) | √(3) | √(0) | √(3) |
| | Trial 2 | √(0) | ×(3) | √(2) | ×(3) | √(0) | ×(3) |
| | Trial 3 | √(0) | ×(3) | √(3) | ×(3) | √(1) | ×(3) |
| Intermediate description | Trial 1 | √(0) | √(1) | √(0) | √(1) | √(3) | √(0) |
| | Trial 2 | √(0) | ×(3) | √(0) | ×(3) | √(1) | ×(3) |
| | Trial 3 | √(3) | √(1) | √(1) | √(3) | √(1) | √(3) |
| Sophisticated description | Trial 1 | √(0) | √(2) | √(0) | √(2) | √(2) | √(0) |
| | Trial 2 | √(0) | √(0) | √(0) | √(0) | √(1) | √(2) |
| | Trial 3 | √(0) | √(0) | √(1) | √(1) | √(1) | √(2) |

Table.A2 Test results of ChatGPT 4.0

| Prompt type | Repeated trials | Model assessment | | | | Code assessment | |
|---|---|---|---|---|---|---|---|
| | | Objective correctness | Constraints correctness | Constraints completeness | Total model Correctness | Error-free execution | Decision verification |
| Simple description | Trial 1 | √(0) | √(1) | √(1) | √(1) | √(1) | √(0) |
| | Trial 2 | √(1) | √(0) | √(1) | √(1) | √(0) | √(0) |
| | Trial 3 | √(0) | √(0) | √(1) | √(1) | √(1) | √(0) |
| Intermediate description | Trial 1 | √(1) | √(0) | √(1) | √(1) | √(0) | √(0) |
| | Trial 2 | √(1) | √(0) | √(0) | √(1) | √(0) | √(0) |
| | Trial 3 | √(0) | √(1) | √(0) | √(1) | √(0) | √(0) |
| Sophisticated description | Trial 1 | √(0) | √(0) | √(0) | √(0) | √(0) | √(0) |
| | Trial 2 | √(0) | √(0) | √(0) | √(0) | √(0) | √(0) |
| | Trial 3 | √(0) | √(1) | √(0) | √(1) | √(0) | √(0) |

Table.A3 Test results of Claude

| Prompt type | Repeated trials | Model assessment | | | | Code assessment | |
|---|---|---|---|---|---|---|---|
| | | Objective correctness | Constraints correctness | Constraints completeness | Total model Correctness | Error-free execution | Decision verification |
| Simple description | Trial 1 | √(1) | ×(3) | √(2) | ×(3) | √(2) | ×(3) |
| | Trial 2 | ×(3) | ×(3) | √(2) | ×(3) | √(0) | ×(3) |
| | Trial 3 | √(0) | ×(3) | √(2) | ×(3) | √(1) | ×(3) |
| Intermediate description | Trial 1 | √(2) | √(0) | √(0) | √(2) | √(0) | √(0) |
| | Trial 2 | √(3) | √(0) | √(0) | √(3) | √(0) | √(0) |
| | Trial 3 | √(1) | √(3) | √(0) | √(3) | √(1) | √(0) |
| Sophisticated description | Trial 1 | √(0) | √(0) | √(0) | √(0) | √(1) | √(0) |
| | Trial 2 | √(0) | √(0) | √(0) | √(0) | √(0) | √(0) |
| | Trial 3 | √(0) | √(0) | √(0) | √(0) | √(0) | √(0) |

Table.A4 Test results of Google Bard

| Prompt type | Repeated trials | Model assessment | | | | Code assessment | |
|---|---|---|---|---|---|---|---|
| | | Objective correctness | Constraints correctness | Constraints completeness | Total model Correctness | Error-free execution | Decision verification |
| Simple description | Trial 1 | ×(3) | ×(3) | √(3) | ×(3) | ×(3) | ×(3) |
| | Trial 2 | ×(3) | ×(3) | √(3) | ×(3) | ×(3) | ×(3) |
| | Trial 3 | ×(3) | ×(3) | √(3) | ×(3) | ×(3) | ×(3) |
| Intermediate description | Trial 1 | ×(3) | ×(3) | √(2) | ×(3) | ×(3) | ×(3) |
| | Trial 2 | ×(3) | ×(3) | √(1) | ×(3) | ×(3) | ×(3) |
| | Trial 3 | ×(3) | ×(3) | √(1) | ×(3) | ×(3) | ×(3) |
| Sophisticated description | Trial 1 | √(0) | ×(3) | √(0) | ×(3) | ×(3) | ×(3) |
| | Trial 2 | √(0) | ×(3) | √(0) | ×(3) | ×(3) | ×(3) |
| | Trial 3 | ×(3) | ×(3) | √(0) | ×(3) | ×(3) | ×(3) |

**3. Links to the dialogue process for accelerating large-scale power system dispatching are as follow:**
https://chat.openai.com/share/ae2257c3-e8ed-427c-9885-e18857e74899
https://chat.openai.com/share/b629d6a5-b6f3-4e4b-856f-e022e2ba8aae
https://chat.openai.com/share/a98e0f3b-239c-4001-831f-5135561f2b56

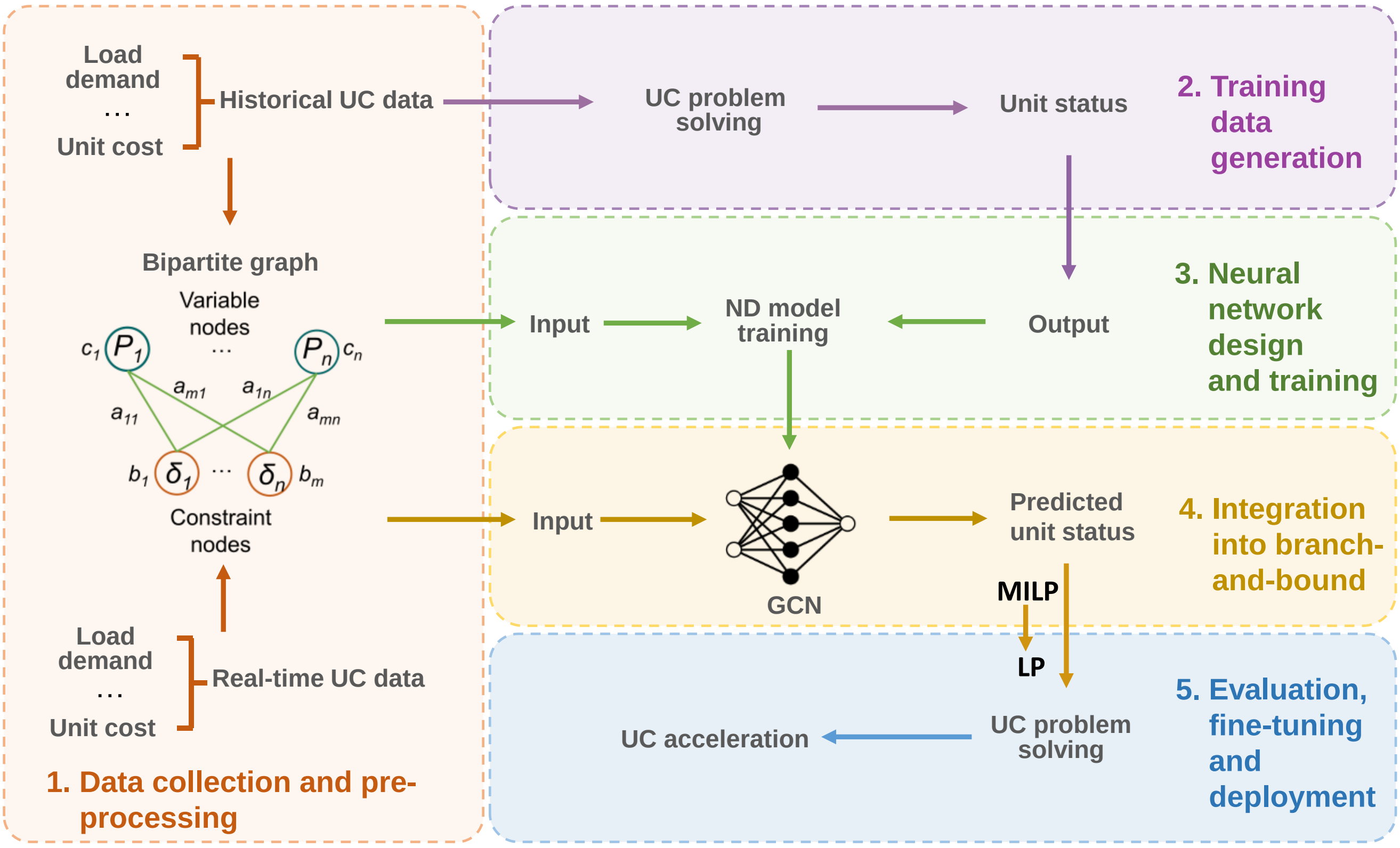

**Extended Data Fig. 1 | Implementation process of ND based on the route designed by ChatGPT4.0.** During the 'Data collection and pre-processing' phase, historical UC load demands and unit costs are amalgamated into a bipartite graph format to facilitate the representation of UC instances. The 'Training data generation' phase determines the on/off states of these units from historical instances through solvers. In the 'Neural network design and training' phase, the integrated bipartite graphs of UC instances and their historical on/off states serve as the training data for the Graph Convolutional Networks (GCN). In 'Integration into branch-and-bound', a well-trained GCN initially predicts the unit's on/off status, subsequently constraining it. This transition recasts the UC issue into a linear programming problem with only continuous variables. In 'Evaluation and fine-tuning and deployment', compared to the traditional branch-and-bound methods, this approach expedites UC solutions.